\documentclass[referee]{raa}            % referee version: for submission

\usepackage{graphicx,times}             %for PS/EPS graphics inclusion, new
\usepackage{natbib}
\usepackage{amssymb,amsmath}
\usepackage{multirow}
\bibpunct{(}{)}{;}{a}{}{,}

\usepackage[a4paper=true,pagebackref=true]{hyperref}
\hypersetup{colorlinks = true, linkcolor = green, anchorcolor = red, citecolor = blue, filecolor = red, pagecolor = red, urlcolor = red}

\begin{document}

   \title{An approximately analytical solution method for the cable-driven parallel robot in FAST
%\,$^*$
%\footnotetext{$*$ Supported by the National Natural Science Foundation of China.}
}
%   \subtitle{I. Place Your Subtitle Here}

   \volnopage{Vol.0 (20xx) No.0, 000--000}      %%preserved for Editor. DOn't remove!
   \setcounter{page}{1}          %%starting page, preserved for Editor. DOn't remove!

   \author{Jia-Ning Yin
      \inst{1,2}
   \and Peng Jiang
      \inst{1}
   \and Rui Yao
      \inst{1}
   }
%% Here is an example of three authors come from different institutes.
%% For single author or all the authors from an institute, use "\inst{}" only

   \institute{National Astronomical Observatories, Chinese Academy of Sciences,
             Beijing 100101, China; {\it pjiang@nao.cas.cn, ryao@nao.cas.cn}\\
%% Please give the E-mail address of the author, to whom future correspondence and
%% offprint requests will be sent.
        \and
             School of Astronomy and Space Science, University of Chinese Academy of Sciences,
             Beijing 100049, China; {\it yinjianing17@mails.ucas.ac.cn}\\           
\vs\no
   {\small Received~~20xx month day; accepted~~20xx~~month day}}

\abstract{FAST is the largest single-dish aperture telescope 
	with a cable-driven parallel robot introduced to achieve the highest sensitivity in the world. 
	However, to realize the high-precision, mechanical equations of such robot are always complicated, 
	so that it is difficult to achieve real-time control by the traditional iterative method. 
	In this regard, this paper proposes an approximately analytical solution method, 
	which uses the approximately linear relationship between the main parameters of FAST to bypass some iterations.  
	With the coefficients of the relationship extracted, static or quasi-static mechanical equations can be analytically solved. 
	In this paper's example, this method saves at least 90\% of the calculating time 
	and the calculated values are consistent with the experimental data. 
	With such huge efficiency improvements, real-time and high-precision control of FAST will no longer be a difficult work. 
	Besides, all the work in this paper is expected to be used in the FAST.
}

   \authorrunning{J.N. Yin, P. Jiang \& R. Yao}            %author_head in even pages
   \titlerunning{AAS method for the CDPR in FAST}  % title_head in odd pages

   \maketitle
%% The author head (on even pages) and the title head (on odd pages) will be
%% automatically extracted from \author{} and \title{}. Whenever the title is too long,
%% you will be asked to supply a shorter one by inserting either \authorrunning{} or
%% \titlerunning{} before \maketitle. Anyway, you can specify your own heads.
%%
%%
%% Note: In the following text body of your manuscript, please note several differences from
%%       other major journals:
%% (1) \subsection{Please Capitalize the First Letter of Each Notional Word in Subsection Title}
%% (2) Please Capitalize the First Letter of Each Notional Word in all tables' captions

%
%________________________________________________ sections below
%
\section{Introduction}           %% first-level sections will be auto-capitalized
\label{sect:intro}

FAST is the largest single-dish aperture telescope with the highest sensitivity in the world. 
To achieve it, the cable-driven parallel robot (CDPR) was introduced (\citealt{Tang+Yao+2011}), 
which mainly composed of some cables and the end effector connected to the cables. 
As shown in Figure 1, there are 6 cables connected to the feed cabin in the center. 
The translation and rotation of the feed cabin can be controlled by adjusting the length of the cables. 
This kind of robot not only has the advantages of high precision, high speed and high load, but also has a large working space. Therefore, CDRP is FAST's perfect solution to solve the wide range movement of the feed cabin. 
Besides, CDPR is not only used in FAST, but also in many fields. 
For example, \cite{Kawamura+etal+1995} developed a robot for transport, called Falcon; 
\cite{Abbasnejad+etal+2016} designed a robot for gait rehabilitation; 
\cite{Bruckmann+etal+2012} also invented a robot related to storage technology, and so on.

According to the study by \cite{Ming+Higuchi+1994}, CDPRs can be divided into three categories. 
Firstly, it is assumed that the number of cables is $m$ and the degree of freedom of the end effector is $n$. 
Thus, if $m=n+1$, the system dynamics equation has a definite solution, 
so it is called completely restrained positioning mechanism (CRPM). 
if $m>n+1$, the driving forces of the cables are redundant, and the system dynamic equation has no definite solution, 
which is called redundantly restrained positioning mechanism (RRPM). 
If $m<n+1$, the system constraints are insufficient. It is called incompletely rested positioning mechanism (IRPM), 
which need to rely on external forces to maintain the stability of the mechanism. 
As mentioned above, FAST’s feed cabin is controlled by 6 cables.
And the degree of freedom of the feed cabin is 6. 
Obviously, FAST is IRPM, so it needs to be stabilized by gravity.

The research on the statics or dynamics of CDPRs must focus on the theoretical model of the cable, 
which determines the mechanical properties of the entire system. 
In this regard, many scholars use the straight line as the cable model 
(\citealt{Cui+etal+2019,Gonzalez-Rodriguez+etal+2017,Vafaei+etal+2017,Kawamura+etal+2000,Khosravi+Taghirad+2013}), 
only considering the elastic deformation of the axial direction of the cables and ignoring the weight influence of the cables. 
This model has analytical expressions and can be solved fast, so it is ideal for CDPRs with a small span. 
However, as for the case of FAST with a large span, the weight influence of the cables cannot be ignored, 
and the cable forces are extremely sensitive to the length of the cables. 
Obviously, the straight-line model is no longer applicable. 
In this regard, other scholars (\citealt{Kozak+etal+2006,Merlet+2019,Yuan+etal+2015}) 
introduced the catenary model derived by \cite{Irvine+1981}. 
It has been verified by \cite{Riehl+etal+2010} that the catenary model has high accuracy. 
However, at the same time, the catenary model needs to be solved by iteration due to its complex nonlinear nature. 
In order to optimize the iteration time, 
\cite{Merlet+2019} proposed to simplify the iteration by changing variables based on the catenary model. 
On the contrary, \cite{Ferravante+etal+2019} abandoned the catenary model and calculated it through finite element method.

However, by now the modeling and solving efficiency of CDPRs has been low, 
because it is inefficient to use the catenary model in real-time control. 
For example, the CDPR control of FAST has to adopt the closed-loop method to save time, 
which is not conducive to increasing its control precision. 
This means that the previous methods cannot achieve real-time and high-precision control at the same time. 
And this paper precisely has a breakthrough at this point.

Based on the static catenary model, with FAST as the research object, 
this paper proposes an approximately analytical solution method. 
This method uses an approximately linear relationship between the main parameters for solution, which is found by numerical analysis. 
After the coefficients of the relationship is extracted, static or quasi-static equations of the CDPR can be simplified and solved analytically, 
which greatly improves the calculation efficiency of FAST.

The approximately analytical solution method for static or quasi-static equations of CDPRs will be introduced in detail. 
For the convenience of description, the following approximately analytical solution method is abbreviated as the AAS method. 
Firstly, the static equations of the FAST cable model will be established, which is  the catenary model, 
and then the relationship between the mechanism parameters and the solution parameters will be analyzed. 
With this relationship, the CDPR's equations can be simplified and solved analytically. 
Finally, there are some example comparisons between the AAS method and the iteration method. The solution accuracy and time of the AAS method are obtained.
Also, there is an experiment, which compares the calculated values of the AAS method 
with the measured values during the actual operation of FAST to test the rationality of the AAS method.

%% Authors can give a citation as 'Michel et al. 1992'.
%% You may also use \cite, \citep and \citet for citation, and use Table~1 or Figure~1
%% and so forth. Using \ref and \label for cross-references of Tables/Figures
%% is a good way in adjusting/adding/removing text, tables or figures.

\section{Coordinate system and parameters}
\label{sect:Coordinate}

This paper takes FAST as the research object, 
which controls the movement and attitude of the central feed cabin by pulling 6 cables through 6 towers. 
It is a typical cable-driven parallel robot (CDPR), as shown in Figure~1.

\begin{figure}
	\centering
	\includegraphics[width=8cm, angle=0]{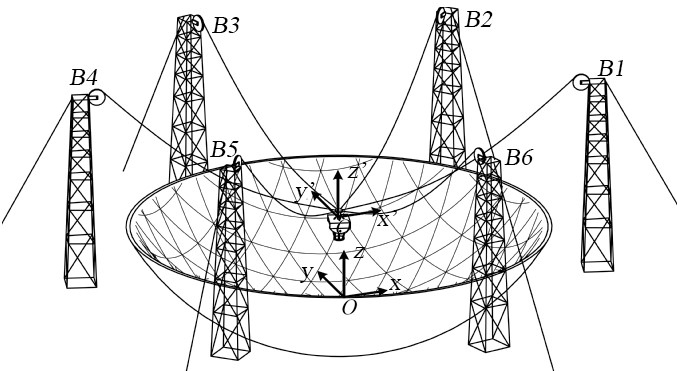}
	\caption{Schematic diagram of the FAST coordinate system.}
	\label{Fig:1}
\end{figure}

Firstly, the global Cartesian coordinate system $O-xyz$ is established. 
The lower vertex of the spherical reflection surface is the origin $O$. 
The direction from the origin $O$ towards the tower $B1$ is the $x$-axis 
and the upward direction perpendicular to the ground is $z$-axis, 
as shown in Figure~1.

Simultaneously, the local Cartesian coordinate system $O'-x'y'z'$ of the feed cabin is also established. 
The center of the anchor points plane of the feed cabin is the origin $O'$. 
The local Cartesian coordinate system is bound to the feed cabin and rotates with the attitude of the feed cabin. 
When the feed cabin is in the center, 
the local Cartesian coordinate system is totally parallel to the global Cartesian coordinate system.

Wherein, the anchor points $A[i]$ of the feed cabin is evenly distributed on the circle with the radius $r_a$. 
The center of the circle just is the origin $O'$ of the local Cartesian coordinate system. 
The 6 towers $B[i]$ are evenly distributed on the circle with the radius $r_b$ and each tower height is $H$. 
Every 2 towers are connected to an anchor point by 2 cables, as shown in Figure~2.

\begin{figure}
	\centering
	\includegraphics[width=\textwidth, angle=0]{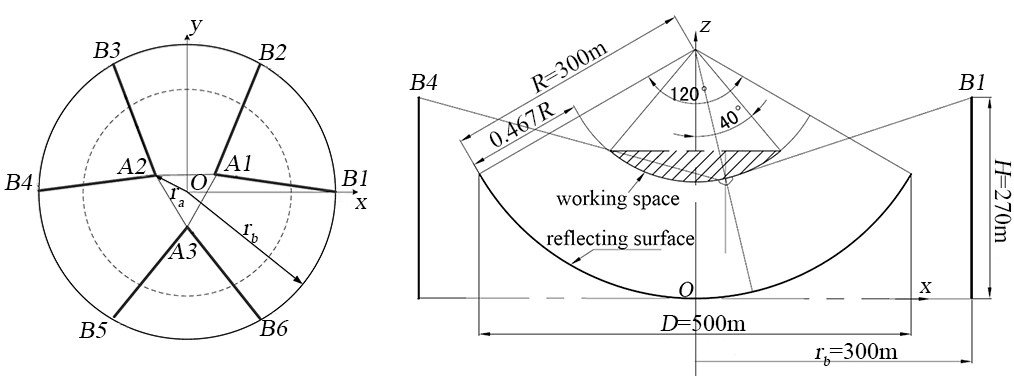}
	\caption{Schematic diagram of the FAST structure.}
	\label{Fig:2}
\end{figure}

Then, the local Cartesian coordinate system $O''-x''y''z''$ of every cable is 
established with the corresponding anchor point $A[i]$ as the origin $O''$. 
For the convenience of calculation, these coordinate systems are 
always required to be parallel with the global Cartesian coordinate system. 
In these coordinate systems, the coordinates of the cable lower and upper end are set to the origin $O''$ and $(X[i],Y[i],Z[i])$, 
respectively, as shown in Figure~3. So, the following geometric relationship can be derived:

\begin{equation}
\begin{bmatrix}
X[i]\\Y[i]\\Z[i]
\end{bmatrix}
=
\begin{bmatrix}
r_b\cos(\pi(i-1)/3)\\r_b\sin(\pi(i-1)/3)\\H
\end{bmatrix}
-(\mathbf{R}\cdot\vec{r_A}[i]+\vec{r_p}) 
\label{eq:1}
\end{equation}

where $\vec{r_A}[i]$ is given by

\begin{equation}
\vec{r_A}[i]
=
\begin{bmatrix}
r_a\cos(\pi/6+2\pi\lfloor(i-1)/2\rfloor/3)\\r_a\sin(\pi/6+2\pi\lfloor(i-1)/2\rfloor/3)\\0
\end{bmatrix}
\label{eq:2}
\end{equation}

Among the formulas above, $[i]$ represents the $i$-th cable corresponding to the $i$-th tower, 
$\mathbf{R}$ is the rotation matrix of the local Cartesian coordinate system $O'-x'y'z'$ of the feed cabin 
relative to the global Cartesian coordinate system $O-xyz$, 
$\vec{r_A}[i]$ is the position vector of the anchor point connected to the $i$-th cable 
in the local Cartesian coordinate system $O'-x'y'z'$ of the feed cabin, 
$\lfloor$ $\rfloor$ is a mathematical symbol, which means rounding down, 
and $\vec{r_p}$ is the position vector of the origin $O'$ of the local Cartesian coordinate system of the feed cabin 
in the global Cartesian coordinate system.

For the specific values of the above and other necessary parameters, please refer to Table~1.

\begin{table}
	\begin{center}
		\caption[]{Specific values of parameters.}
		\label{tab:1}
		\begin{tabular}{clc}
			\hline\noalign{\smallskip}
			Symbol     &Significance                                          &Specific value(unit)                \\
			\hline\noalign{\smallskip}
			$r_a$      & Feed cabin anchor point distribution radius          &$7.5(\mathrm{m})$                   \\
			$r_b$      & Tower distribution radius                            &$300(\mathrm{m})$                   \\
			$H$        & Tower height                                         &$270(\mathrm{m})$                   \\
			$E$        &Cable elastic modulus                                 &$1.6\times10^{11}(\mathrm{Pa})$     \\
			$A$        &Cable cross-sectional area with the cable not stressed&$1.541\times10^{-4}(\mathrm{m^2})$  \\
			$\rho$     &Cable linear density with the cable not stressed      &$11.718(\mathrm{kg/m})$             \\
			$g$        &the acceleration of gravity                           &$-9.8(\mathrm{m/s^2})$              \\
			\multirow{2}{*}{$\vec{r_e}$}&
			Position vector from the origin $O'$ of the feed cabin&
			\multirow{2}{*}{$[0,0,0.5]^{\mathrm{T}}(\mathrm{m})$}\\
					   &local coordinate system to the feed cabin mass center &                                    \\
			$m$        &Feed cabin mass                                       &$30000(\mathrm{kg})$                \\
			$R$        &Reflecting surface radius                             &$300(\mathrm{m})$                   \\
			$D$        &Reflecting surface projection diameter                &$500(\mathrm{m})$                   \\
			\noalign{\smallskip}\hline
		\end{tabular}
	\end{center}
\end{table}

\section{Cable model and derivation}
\label{sect:derivation}

The cable model of this paper is the static catenary model, and the coordinates are shown in Figure~3. 
Because the equations of the 6 cables' model are the same, for the convenience, 
the cable number $i$ is generally not specified in this section unless it is necessary.
Let the forces in the three directions of the cable lower end be $F_x$, $F_y$ and $F_z$, 
respectively, as shown in Figure~4. 

\begin{figure}[h]
	\begin{minipage}[t]{0.495\linewidth}
		\centering
		\includegraphics[width=40mm]{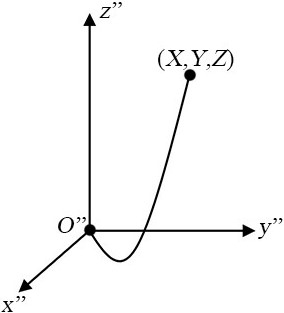}
		\caption{{\small Local coordinate system of the cable.} }
	\end{minipage}%
	\begin{minipage}[t]{0.495\textwidth}
		\centering
		\includegraphics[width=40mm]{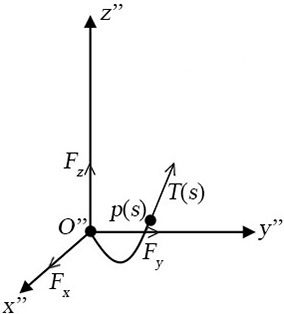}
		\caption{{\small Schematic diagram of the cable force.}}
	\end{minipage}%
	\label{Fig:34}
\end{figure}

Where the length of the cable is $p$, $T$ is the cable force at that point and $s$ is the length of this cable without tension, 
called original length of the cable. 
The length of the cable $p$ and the cable force $T$ are both the functions of the original length $s$. 
In addition, let $\rho$ be the linear density when the cable is not tensioned, and $g$ be the acceleration of gravity. 
Then, according to the equilibrium equation, the following can be obtained:

\begin{equation}
T\frac{\mathrm{d}x}{\mathrm{d}p}+F_{x}=0
\label{eq:3}
\end{equation}
\begin{equation}
T\frac{\mathrm{d}y}{\mathrm{d}p}+F_{y}=0
\label{eq:4}
\end{equation}
\begin{equation}
T\frac{\mathrm{d}z}{\mathrm{d}p}+F_{z}+\rho gs=0
\label{eq:5}
\end{equation}

where $T$ is given by

\begin{equation}
T(s)=\sqrt{F_{x}^{2}+F_{y}^{2}+\left(F_{z}+\rho gs\right)^2}
\label{eq:6}
\end{equation}

Then, according to the elastic equation, there is

\begin{equation}
T(s)=EA\left(\frac{\mathrm{d}p}{\mathrm{d}s}-1\right)
\label{eq:7}
\end{equation}

where $EA$ is the cable elastic modulus multiplied by the cross-sectional area when the cable is not tensioned.

Combined with equation (3-7), the following can be obtained:

\begin{equation}
\frac{\mathrm{d}x}{\mathrm{d}s}=
-\frac{F_x}{EA}\left(1+\frac{EA}{\sqrt{F_{x}^{2}+F_{y}^{2}+\left(F_{z}+\rho gs\right)^2}}\right)  
\label{eq:8}
\end{equation}

\begin{equation}
\frac{\mathrm{d}y}{\mathrm{d}s}=
-\frac{F_y}{EA}\left(1+\frac{EA}{\sqrt{F_{x}^{2}+F_{y}^{2}+\left(F_{z}+\rho gs\right)^2}}\right)  
\label{eq:9}
\end{equation}

\begin{equation}
\frac{\mathrm{d}z}{\mathrm{d}s}=
-\frac{F_z+\rho gs}{EA}\left(1+\frac{EA}{\sqrt{F_{x}^{2}+F_{y}^{2}+\left(F_{z}+\rho gs\right)^2}}\right)  
\label{eq:10}
\end{equation}

According to the boundary conditions $x(0)=0$, $y(0)=0$ and $z(0)=0$, shown in Figure~4, the solutions are equation (11-13).

\begin{equation}
x(s)=-\frac{F_x}{EA}s-\frac{F_x}{\rho g}\left[\sinh^{-1}\left(\frac{F_z+\rho gs}{\sqrt{F_{x}^{2}+F_{y}^{2}}}\right)
		-\sinh^{-1}\left(\frac{F_z}{\sqrt{F_{x}^{2}+F_{y}^{2}}}\right)\right]
\label{eq:11}
\end{equation}

\begin{equation}
y(s)=-\frac{F_y}{EA}s-\frac{F_y}{\rho g}\left[\sinh^{-1}\left(\frac{F_z+\rho gs}{\sqrt{F_{x}^{2}+F_{y}^{2}}}\right)
		-\sinh^{-1}\left(\frac{F_z}{\sqrt{F_{x}^{2}+F_{y}^{2}}}\right)\right]
\label{eq:12}
\end{equation}

\begin{equation}
z(s)=-\frac{F_z}{EA}s-\frac{\rho g}{2EA}s^2-\frac{1}{\rho g}\left[\sqrt{F_{x}^{2}+F_{y}^{2}+\left(F_{z}
		+\rho gs\right)^2}-\sqrt{F_{x}^{2}+F_{y}^{2}+F_{z}^2}\right]
\label{eq:13}
\end{equation}

Let the original length of the whole cable be $s_0$, 
and know that the coordinates of the cable upper end are $(X,Y,Z)$, 
then equation (14-16) can be obtained, where the unknown variables are $F_x$, $F_y$, $F_z$ and $s_0$.

\begin{equation}
X=-\frac{F_x}{EA}s_0-\frac{F_x}{\rho g}\left[\sinh^{-1}\left(\frac{F_z+\rho gs_0}{\sqrt{F_{x}^{2}+F_{y}^{2}}}\right)
-\sinh^{-1}\left(\frac{F_z}{\sqrt{F_{x}^{2}+F_{y}^{2}}}\right)\right]
\label{eq:14}
\end{equation}

\begin{equation}
Y=-\frac{F_y}{EA}s_0-\frac{F_y}{\rho g}\left[\sinh^{-1}\left(\frac{F_z+\rho gs_0}{\sqrt{F_{x}^{2}+F_{y}^{2}}}\right)
-\sinh^{-1}\left(\frac{F_z}{\sqrt{F_{x}^{2}+F_{y}^{2}}}\right)\right]
\label{eq:15}
\end{equation}

\begin{equation}
Z=-\frac{F_z}{EA}s_0-\frac{\rho g}{2EA}s_0^2-\frac{1}{\rho g}\left[\sqrt{F_{x}^{2}+F_{y}^{2}+
	\left(F_{z}+\rho gs_0\right)^2}-\sqrt{F_{x}^{2}+F_{y}^{2}+F_{z}^2}\right]
\label{eq:16}
\end{equation}

In the local coordinate system of the cable, let the resultant force on the $O''-x''y''$ plane be $F_l=\sqrt{F_x^2+F_y^2}$. 
Refer to equation (14) and (15) and the following can be obtained:

\begin{equation}
F_x=\frac{-X}{\sqrt{X^2+Y^2}}F_l 
\label{eq:17}
\end{equation}

\begin{equation}
F_y=\frac{-Y}{\sqrt{X^2+Y^2}}F_l 
\label{eq:18}
\end{equation}

If the cable is straight, the cable length must be $\sqrt{X^2+Y^2+Z^2}$, 
and let $k=\sqrt{X^2+Y^2+Z^2}$. 
Because $EA$ is of a large magnitude, 
the actual original cable length $s_0$ generally does not exceed the interval $[0.95k,1.05k]$. 
Now with FAST as the object, when the feed cabin is at a random position, 
by solving the numerical value of equation (14-16), 
the change trends of $F_l$ and $F_z$ can be obtained with the cable length $s_0$ in the interval above, as shown in Figure~5.

There is an obvious feature in Figure~5. When the cable length shrinks to a certain value, 
the sensitivity of the cable force to the original length $s_0$ of the whole cable rises rapidly, 
but later it quickly remains stable. This is a complex form of function, which leads to difficulties in iteration. 
However, it is observed that the trends $F_l$ and $F_z$ are highly consistent, 
so another figure of $F_z$ on $F_l$ is considered, as shown in Figure~6.

\begin{figure}[h]
	\begin{minipage}[t]{0.495\linewidth}
		\centering
		\includegraphics[width=72mm]{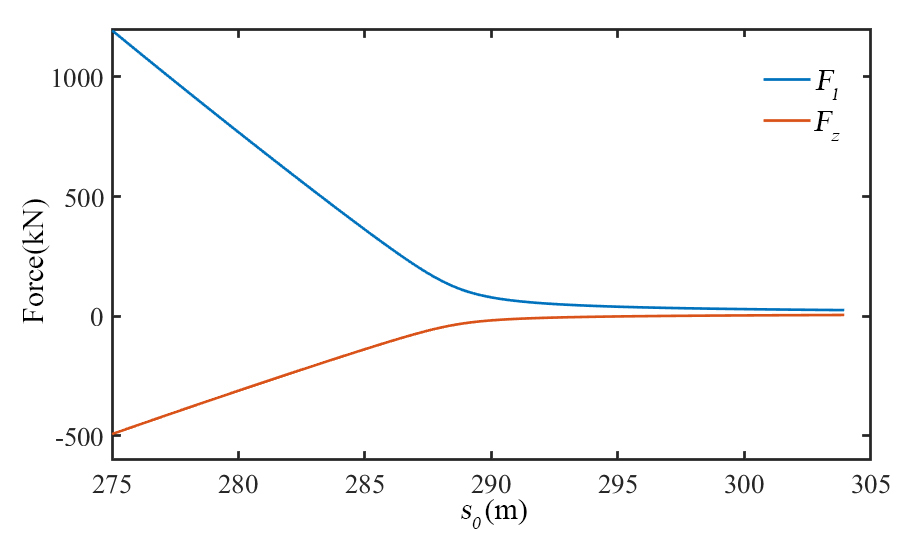}
		\caption{{\small $F_l$ and $F_z$ change with $s_0$.} }
	\end{minipage}%
	\begin{minipage}[t]{0.495\textwidth}
		\centering
		\includegraphics[width=72mm]{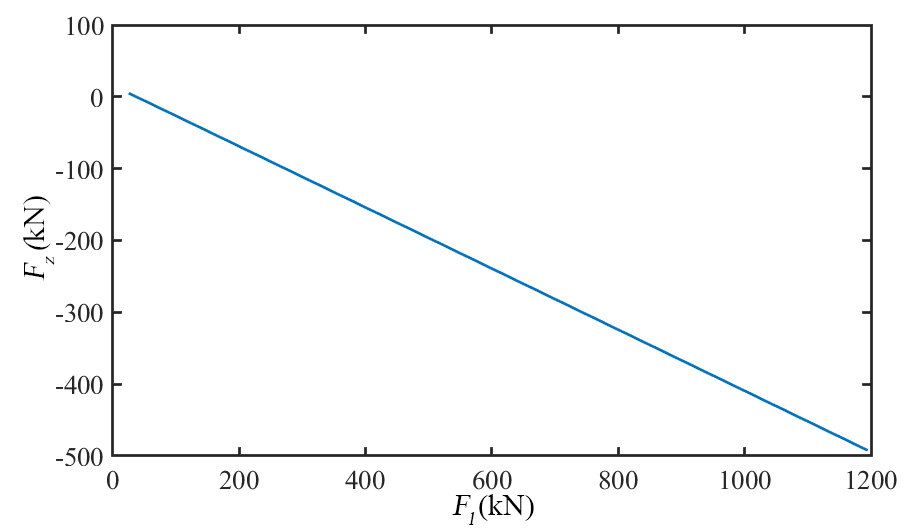}
		\caption{{\small $F_z$ changes with $F_l$.}}
	\end{minipage}%
	\label{Fig:fig56}
\end{figure}

Obviously, $F_z$ has a strong linear relationship with $F_l$, which is much simpler than $F_z$'s case on $s_0$. 
After massive calculation with taking all the position of the feed cabin’s working space in Figure~2, 
it is found that minimum value of the determination coefficient of this linear relationship is 0.999999999750919, 
so the linear relationship can be considered always to exist and  be independent of the original length $s_0$ of the whole cable.

However, it should be noted that the linear relationship is related to 
the spatial structure and physical properties of the research object. 
For each research object, the relationship needs to be verified by numerical calculation in the CDPR's workspace. 
In this paper, FAST has such a good linear relationship.

Therefore, in the actual calculation, it is only necessary to take two kinds of $s_0$ in the equation (14-16). 
For example, $s_0=0.99k$ and $s_0=1.01k$. Then, the linear expression of $F_z$ about $F_l$ can be determined:

\begin{equation}
F_z=aF_l+b
\label{eq:19}
\end{equation}

However, FAST has 6 cables, so there are 6 groups of equation (14-16), 
which means there are a total of 24 unknown variables with only 18 equations. 
So, another 6 equations are needed to solve the equation. Fortunately, the feed cabin balance equations just meet this:

\begin{equation}
-\sum_{i=1}^{6}F_x[i]=0
\label{eq:20}
\end{equation}

\begin{equation}
-\sum_{i=1}^{6}F_y[i]=0
\label{eq:21}
\end{equation}

\begin{equation}
mg-\sum_{i=1}^{6}F_z[i]=0
\label{eq:22}
\end{equation}

\begin{equation}
\sum_{i=1}^{6}\left(-r_y[i]F_z[i]+r_z[i]F_y[i]\right)+mge_y=0
\label{eq:23}
\end{equation}

\begin{equation}
\sum_{i=1}^{6}\left(r_x[i]F_z[i]-r_z[i]F_x[i]\right)+mge_x=0
\label{eq:24}
\end{equation}

\begin{equation}
\sum_{i=1}^{6}\left(-r_x[i]F_y[i]+r_y[i]F_x[i]\right)=0
\label{eq:25}
\end{equation}

where $[i]$ represents the $i$-th cable and $m$ is the feed cabin mass, 
and $[r_x[i],r_y[i],r_z[i]]^\mathrm{T}=\mathbf{R}\cdot\vec{r_A}[i]$.
$e_x$ and $e_y$ are the projection distances of the position $\vec{r_e}$ shown in Table 1, 
respectively in the $x$-axis direction and the $y$-axis direction.

Substitute equation (17-19) into equation (20-25), which can be reduced to the following matrix form:

\begin{equation}
\mathbf{A}\cdot\vec{F_l}=\vec{B}
\label{eq:26}
\end{equation}

where $\mathbf{A}$, $\vec{F_l}$ and $\vec{B}$ are given by equation (27-29).

\begin{equation}
\mathbf{A}=\begin{bmatrix}  
\frac{X[1]}{\sqrt{X[1]^2+Y[1]^2}} & \cdots & \frac{X[6]}{\sqrt{X[6]^2+Y[6]^2}}\\  
\frac{Y[1]}{\sqrt{X[1]^2+Y[1]^2}} & \cdots & \frac{Y[6]}{\sqrt{X[6]^2+Y[6]^2}}\\
a[1]                              & \cdots & a[6]                             \\  
\frac{r_z[1]Y[1]+r_y[1]a[1]\sqrt{X[1]^2+Y[1]^2}}{\sqrt{X[1]^2+Y[1]^2}}&\cdots
&\frac{r_z[6]Y[6]+r_y[6]a[6]\sqrt{X[6]^2+Y[6]^2}}{\sqrt{X[6]^2+Y[6]^2}}\\
\frac{r_z[1]X[1]+r_x[1]a[1]\sqrt{X[1]^2+Y[1]^2}}{\sqrt{X[1]^2+Y[1]^2}}&\cdots
&\frac{r_z[6]X[6]+r_x[6]a[6]\sqrt{X[6]^2+Y[6]^2}}{\sqrt{X[6]^2+Y[6]^2}}\\
\frac{r_y[1]X[1]-r_x[1]Y[1]}{\sqrt{X[1]^2+Y[1]^2}}&\cdots&
\frac{r_y[6]X[6]-r_x[6]Y[6]}{\sqrt{X[6]^2+Y[6]^2}}\\
\end{bmatrix}
\label{eq:27}
\end{equation}

\begin{equation}
\vec{F_l}=\begin{bmatrix}  
F_l[1]&F_l[2]& F_l[3]&F_l[4]&F_l[5]&F_l[6]\\  
\end{bmatrix}^\mathrm{T} 
\label{eq:28}
\end{equation}

\begin{equation} 
\vec{B}=\begin{bmatrix}  
0&0&(mg-\sum_{i=1}^{6}b[i])&(mge_y-\sum_{i=1}^{6}r_y[i]b[i])&(mge_x-\sum_{i=1}^{6}r_x[i]b[i])&0\\  
\end{bmatrix}^\mathrm{T} 
\label{eq:29}
\end{equation}

Therefore, it is easy to get the resultant force $F_l$ of each cable on the respective $O''-x''y''$ plane, 
which is also on the global plane $O-xy$, 
because the local Cartesian coordinate system $O''-x''y''z''$ of each cable 
is parallel with the global Cartesian coordinate system $O-xyz$.

\begin{equation} 
\vec{F_l}=\mathbf{A}^{-1}\cdot\vec{B}
\label{eq:30}
\end{equation}

Then according to the equation (17-19), 
the forces $F_x$, $F_y$ and $F_z$ of the lower end of each cable can be obtained.

It can be seen that the form of equation (26) is very similar to the straight-line model’s. 
The difference is in the matrix $\mathbf{A}$ and the array $\vec{B}$. 
New parameters $a[i]$ and $b[i]$ are introduced, so that the expression not only corresponds to the geometric relationship, 
but also the mechanical parameters of the cable and the attitude of the feed cabin. 
In a sense, $a[i]$ and $b[i]$ are equivalent to the correction parameters used to correct the error 
between the linear model and the catenary model, 
which depend on the mechanical and geometric properties of the entire system.

So far, the process of solving the static or quasi-static equations of CDPRs 
by the approximately analytical solution method (AAS) has been very clear, see Figure~7 for details.

\begin{figure}
	\centering
	\includegraphics[width=8cm, angle=0]{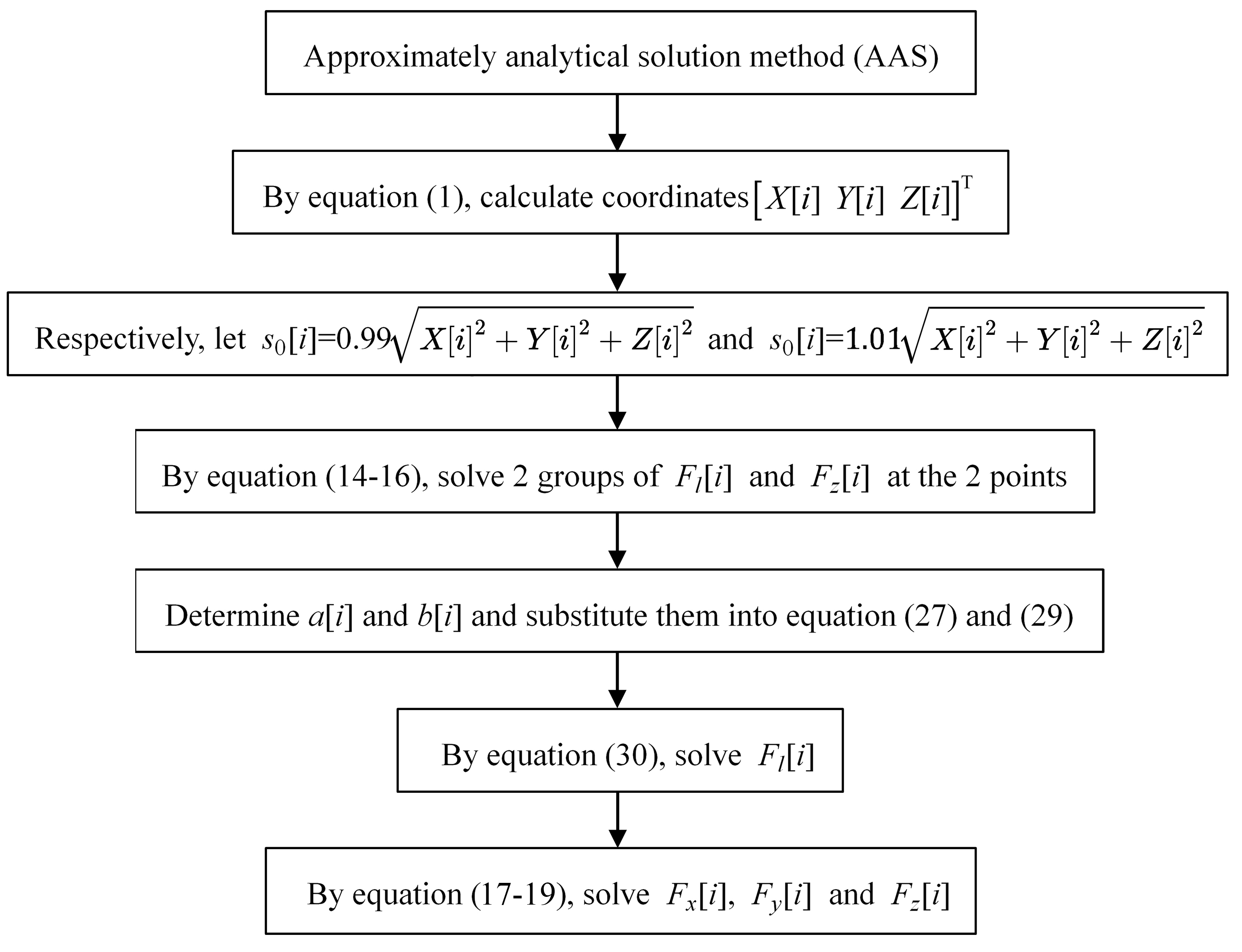}
	\caption{Process of approximately analytical solution method.}
	\label{Fig:7}
\end{figure}

Obviously, the process can solve all the required parameters just in one loop. 
Compared to the traditional iterative operation, there is no step of loop calculation and selecting step size. 
For this reason, AAS method can greatly improve the static solution speed of the CDPR in FAST.

\section{Examples and comparison}
\label{sect:Comparison}

In this section, a comparison between the approximately analytical solution method and the traditional iterative method will be shown, 
based on MAPLE programming. Under the condition of the same feed cabin trajectory, 
the same static or quasi-static equations of FAST's CDPR are solved by the two methods respectively. 
Finally, the cable force values of the lower ends of the six cables and the time required for the solution will be compared.

\begin{figure}
	\centering
	\includegraphics[width=8cm, angle=0]{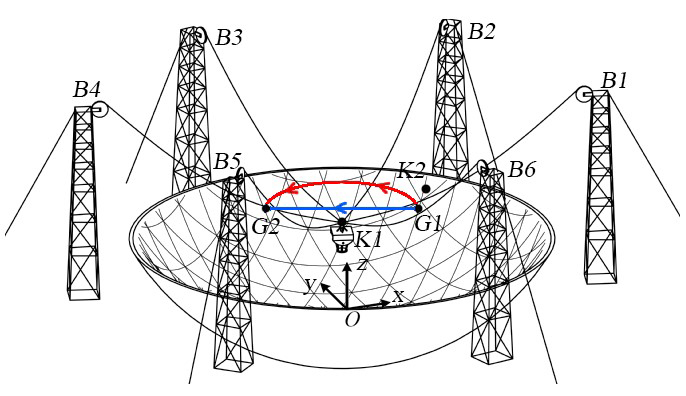}
	\caption{Schematic diagram of the feed cabin trajectory.}
	\label{Fig:8}
\end{figure}

(1) The feed cabin is hovering at the lower vertex $(0,0,140\mathrm{m})$ of the working area, 
which means the feed cabin remains stationary at point $K1$ in Figure~8. 
Because the feed cabin is in the center, according to the principle of symmetry, 
the six cables should be subjected to the same force. Table~2 shows the calculation results.

In the case of high symmetry, the relative error between AAS and iteration method is very small. 
It can be considered that the two methods have similar accuracy, 
but the time cost of AAS is obviously much less than the iteration method.

\begin{table}
	\begin{center}
		\caption[]{Comparison with the feed cabin hovering at the lower vertex of the workspace.}
		\label{tab:2}
		\begin{tabular}{cccccccc}
			\hline\noalign{\smallskip}
			\multirow{2}{*}{Method} & \multicolumn{6}{c}{Cable force at the lower end/kN} & \multirow{2}{*}{Time cost/s} \\ 
			\cline{2-7}\noalign{\smallskip}
			& 1        & 2        & 3        & 4        & 5        & 6        &       \\ 
			\hline\noalign{\smallskip}
			Iteration      & 159.1762 & 159.1670 & 159.1587 & 159.1587 & 159.1670 & 159.1762 & 4.984 \\
			AAS            & 158.9899 & 158.9808 & 158.9725 & 158.9725 & 158.9808 & 158.9899 & 0.938 \\
			Relative error & 0.1170\% & 0.1170\% & 0.1170\% & 0.1170\% & 0.1170\% & 0.1170\% & ——    \\
			\noalign{\smallskip}\hline                       
		\end{tabular}
	\end{center}
\end{table}

(2) The feed cabin is hovering at a point that is not specific in the working area, 
such as the point $K2$ $(42.65\mathrm{m},33.73\mathrm{m},149.52\mathrm{m})$, shown in Figure~8. 
This point is closer to the $B1$, $B2$, $B3$ and $B6$ towers, so the cable tension of the four towers should be larger. 
Table~3 shows the calculation results.

In the case of no special position, the relative error of AAS with the iteration is still very small. 
It can be considered that the two methods have similar accuracy. And AAS is obviously much faster than the iteration.

\begin{table}
	\begin{center}
		\caption[]{Comparison with the feed cabin hovering at a point that is not specific.}
		\label{tab:3}
		\begin{tabular}{cccccccc}
			\hline\noalign{\smallskip}
			\multirow{2}{*}{Method} & \multicolumn{6}{c}{Cable force at the lower end/kN} & \multirow{2}{*}{Time cost/s} \\ 
			\cline{2-7}\noalign{\smallskip}
			& 1        & 2        & 3        & 4        & 5        & 6        &       \\ 
			\hline\noalign{\smallskip}
			Iteration      & 181.5009 & 190.7462 & 200.7024 & 119.5531 & 119.8599 & 191.0149 & 4.594 \\
			AAS            & 181.3146 & 190.5478 & 200.6720 & 119.3861 & 119.6912 & 190.9943 & 0.969 \\
			Relative error & 0.1026\% & 0.1040\% & 0.0151\% & 0.1397\% & 0.1407\% & 0.0108\% & ——    \\
			\noalign{\smallskip}\hline                       
		\end{tabular}
	\end{center}
\end{table}

(3) The feed cabin slowly moves in a straight path 
from $G1$ $(50\mathrm{m},0,150\mathrm{m})$ to $G2$ $(-50\mathrm{m},0,150\mathrm{m})$, 
as shown in the blue line in Figure~8. 
Because the feed cabin’s movement is very slow, it can be considered that the system is quasi-static during the whole process. 
In the solution, the trajectory is evenly divided into 101 nodes. 
The static equations of the CDPR of each node are solved by the two methods. 
Figure~9 shows the change of the forces of the six cables in the whole process, 
and the abscissa is the distance traveled by the feed cabin.

The cable forces solved by the two methods are almost identical. 
Table~4 lists the maximum relative error of each cable force during the calculation process and the solution time. 
With the same solution accuracy, AAS takes much less time than the iteration, 
which is important for FAST to achieve real-time control and improve accuracy.

\begin{table}
	\begin{center}
		\caption[]{Comparison with the feed cabin moving slowly in a straight line.}
		\label{tab:4}
		\begin{tabular}{cccccccc}
			\hline\noalign{\smallskip}
			\multicolumn{6}{c}{Maximum relative error of cable force at the lower end} & \multicolumn{2}{c}{Time cost/s} \\ 
			\hline\noalign{\smallskip}
			1 & 		  2 &         3 &         4 &         5 &         6 &   AAS &Iteration\\ 
			\hline\noalign{\smallskip}
			0.1140\%	&0.1020\%	&0.1020\%	&0.1140\%	&0.1197\%	&0.1197\%	&20.844	&343.203  \\
			\noalign{\smallskip}\hline                       
		\end{tabular}
	\end{center}
\end{table}

\begin{figure}[h]
	\begin{minipage}[t]{0.495\linewidth}
		\centering
		\includegraphics[width=70mm]{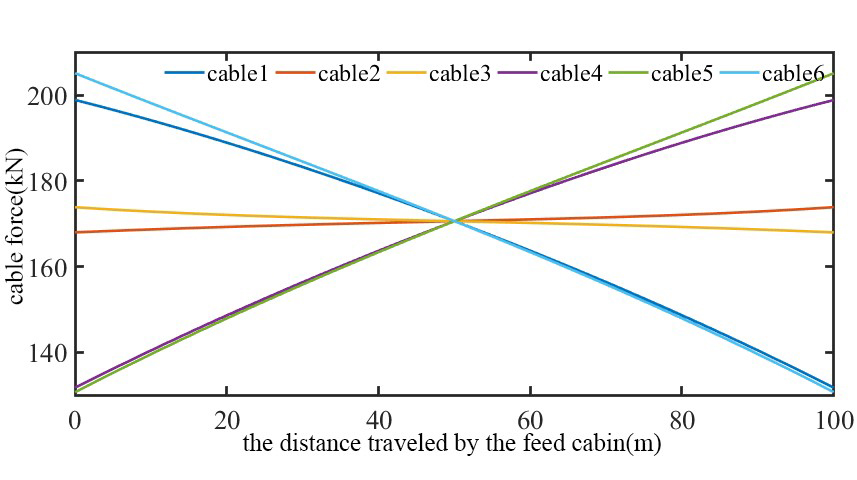}\\
		{\tiny (a) Cable forces solved by AAS}
	\end{minipage}%
	\begin{minipage}[t]{0.495\textwidth}
		\centering
		\includegraphics[width=70mm]{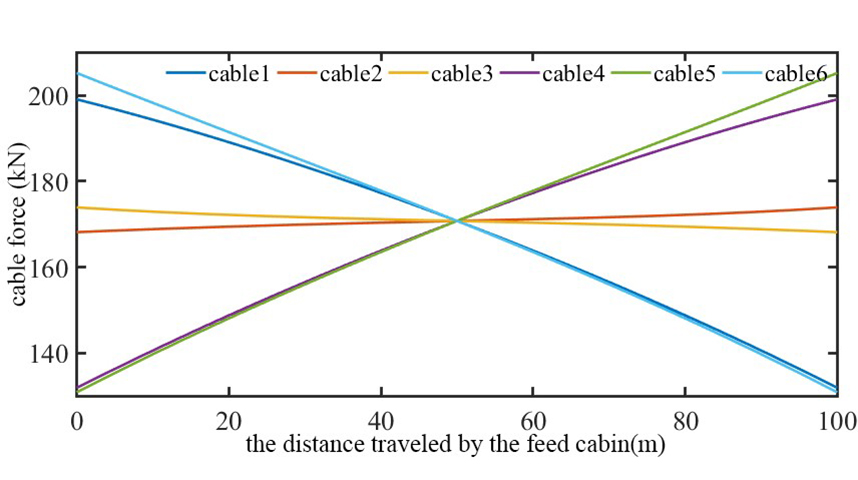}\\
		{\tiny (b) Cable forces solved by the iterative}
	\end{minipage}%
	\caption{\small Comparison with the feed cabin moving slowly in a straight line.}
	\label{Fig:9}
\end{figure}

(4) The feed cabin slowly moves in a circular path 
with the point $(0,0,150\mathrm{m})$ as the center and $50\mathrm{m}$ as the radius, 
keeping the height unchanged from $G1$ $(50\mathrm{m},0,150\mathrm{m})$ to $G2$ $(-50\mathrm{m},0,150\mathrm{m})$, 
see the red line in Figure~8. As in the previous case, it can be considered that the entire system is quasi-static, 
and the trajectory is divided into 101 nodes to solve one by one. 
Figure~10 shows the change of the forces of the six cables in the whole process, 
and the abscissa is the distance traveled by the feed cabin.

Like the case of the straight line, the cable forces solved by the two methods are almost identical. 
Table~5 lists the maximum relative error of each cable force during the calculation process and the solution time. 
With the same solution accuracy, AAS is still much faster than the iteration.

\begin{table}
	\begin{center}
		\caption[]{Comparison with the feed cabin moving slowly in a circular path.}
		\label{tab:5}
		\begin{tabular}{cccccccc}
			\hline\noalign{\smallskip}
			\multicolumn{6}{c}{Maximum relative error of cable force at the lower end} & \multicolumn{2}{c}{Time cost/s} \\ 
			\hline\noalign{\smallskip}
			1 & 		  2 &         3 &         4 &         5 &         6 &   AAS &Iteration\\ 
			\hline\noalign{\smallskip}
			0.1308\%	&0.1016\%	&0.1017\%	&0.1308\%	&0.1308\%	&0.1308\%	&22.750	&372.000\\
			\noalign{\smallskip}\hline                       
		\end{tabular}
	\end{center}
\end{table}

\begin{figure}[h]
	\begin{minipage}[t]{0.495\linewidth}
		\centering
		\includegraphics[width=70mm]{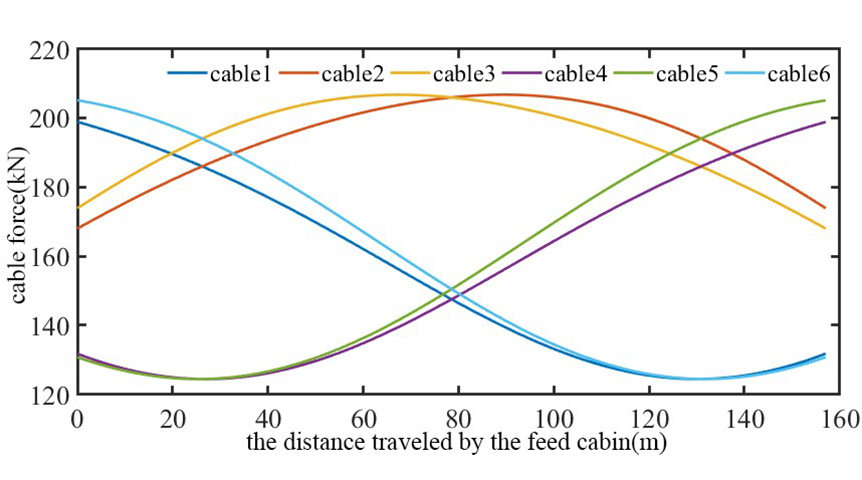}\\
		{\tiny (a) Cable forces solved by AAS}
	\end{minipage}%
	\begin{minipage}[t]{0.495\textwidth}
		\centering
		\includegraphics[width=70mm]{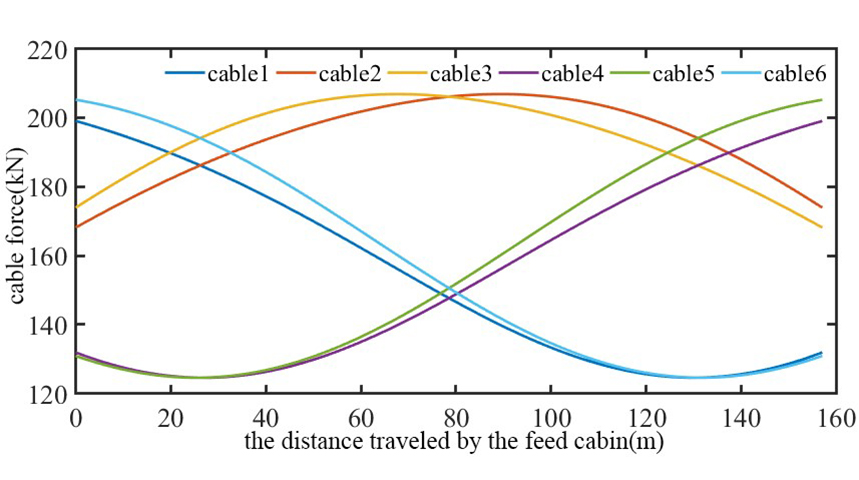}\\
		{\tiny (b) Cable forces solved by the iterative}
	\end{minipage}%
	\caption{\small Comparison with the feed cabin moving slowly in a circular path.}
	\label{Fig:10}
\end{figure}

It can be seen from the comparison above that the calculation accuracy of the AAS method 
for solving FAST's cable forces is comparable to the iterative method, and the calculation time is greatly reduced. 
However, because the applied catenary model is a static model, this method is best applied to static or quasi-static situations. 
Whether this theory can be applied to dynamic calculations requires in-depth analysis combined with the actual model and further research.

The following is a comparison between the AAS method and the method currently used in FAST. 
By letting the feed cabin run the same trajectory, 
the cable forces calculated by the AAS method are compared with the cable forces 
measured by the sensors when FAST is actually controlled. 
These sensors are respectively  installed on 6 cables as close as possible to the anchor points $A[i]$ shown in Figure~2.
The trajectory is shown in Figure~11, with $(0,0,156.73\mathrm{m})$ as the center, $71.11\mathrm{m}$ as the radius, 
and making a full circle from $G3 (0,71,11\mathrm{m},156.73\mathrm{m})$ while maintaining the same height.
It should be noted that the running process is slow and the system can be considered as quasi-static.

\begin{figure}
	\centering
	\includegraphics[width=8cm, angle=0]{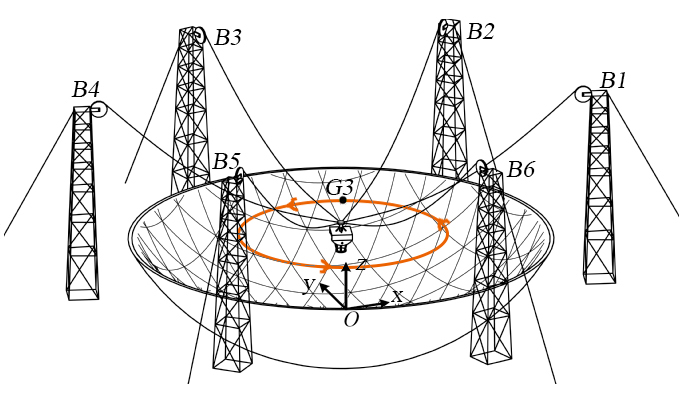}
	\caption{Schematic diagram of the feed cabin trajectory.}
	\label{Fig:11}
\end{figure}

\begin{table}
	\begin{center}
		\caption[]{Root mean square errors between AAS method and the real-time control.}
		\label{tab:6}
		\begin{tabular}{cccccccc}
			\hline\noalign{\smallskip}
			\multirow{2}{*}{Method} & \multicolumn{6}{c}{Root mean square error of cable force/kN}\\ 
			\cline{2-7}\noalign{\smallskip}
			& 1        & 2        & 3        & 4        & 5        & 6        \\ 
			\hline\noalign{\smallskip}
			AAS      & 22.554 & 17.517 & 20.078 & 18.687 & 17.167 & 18.190 \\
			AAS with actual attitudes& 23.942 & 14.226 & 19.531 & 14.210 & 17.534 & 16.058 \\
			AAS with actual attitudes and corrected coordinates & 20.060 & 12.683 & 16.422 & 10.018 & 16.457 & 12.053 \\
			\noalign{\smallskip}\hline                       
		\end{tabular}
	\end{center}
\end{table}

Figure~12(a) shows the theoretically calculated cable forces as the feed cabin moves under this trajectory, 
while Figure~12(b) shows the actual cable forces measured during real-time control. 
And the root mean square errors between them are shown in Table~6. 
It can be seen that the theoretical and experimental numerical trends are consistent, 
but there are still considerable discrepancies. 
Considering that the attitude change of the feed cabin has a huge influence on the cable force, 
it is necessary to use the feed cabin attitude measured in real-time control when using the AAS method for calculation.

\begin{figure}[h]
	\begin{minipage}[t]{0.495\linewidth}
		\centering
		\includegraphics[width=70mm]{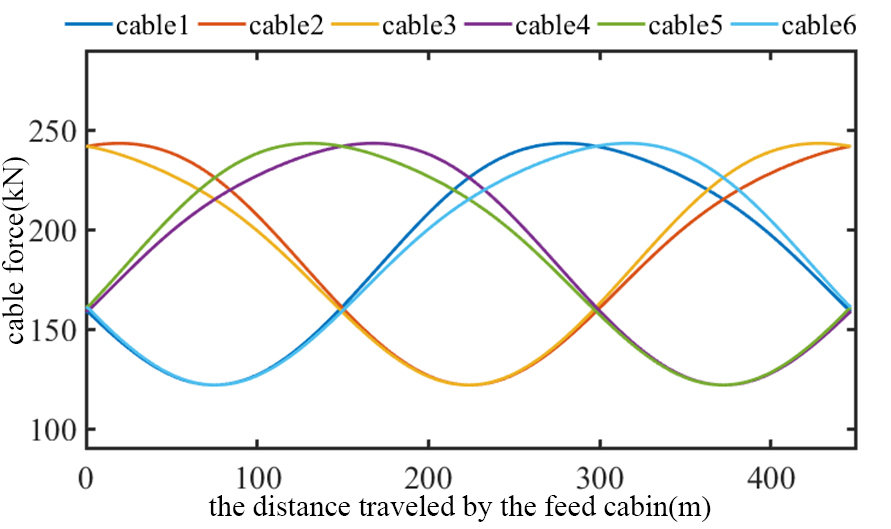}\\
		{\tiny (a) Cable forces solved by AAS}
	\end{minipage}%
	\begin{minipage}[t]{0.495\textwidth}
		\centering
		\includegraphics[width=70mm]{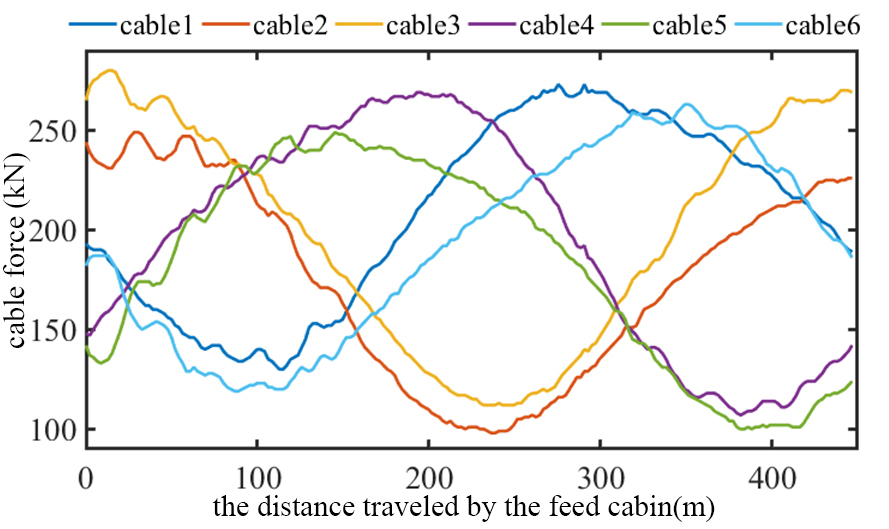}\\
		{\tiny (b) Cable forces measured during real-time control}
	\end{minipage}%
	\caption{\small Comparison between the AAS method and the real-time control.}
	\label{Fig:12}
\end{figure}

Figure~13(a) shows the cable forces calculated by the AAS method after considering the measured attitude of the feed cabin, 
while Figure~13(b) shows the relative errors between these calculated cable forces 
and the actual measured cable forces in Figure~12(b). 
And the root mean square errors between them are shown in Table~6. 
It can be seen that the theoretical and experimental numerical trends are more consistent, 
but there are still some deviations. 
After careful inspection, it was found that the mass center coordinates of the feed cabin had a large deviation. 
After iterative calculation, it is finally determined that the mass center is near $(0.22\mathrm{m},0.11\mathrm{m},0.5\mathrm{m})$, 
which is far away from the theoretical coordinates $(0,0,0.5\mathrm{m})$. 
This also leads to larger deviations of cable forces. 
Therefore, it is necessary to correct the mass center coordinates to recalculate the cable forces by the AAS method.

\begin{figure}[h]
	\begin{minipage}[t]{0.495\linewidth}
		\centering
		\includegraphics[width=70mm]{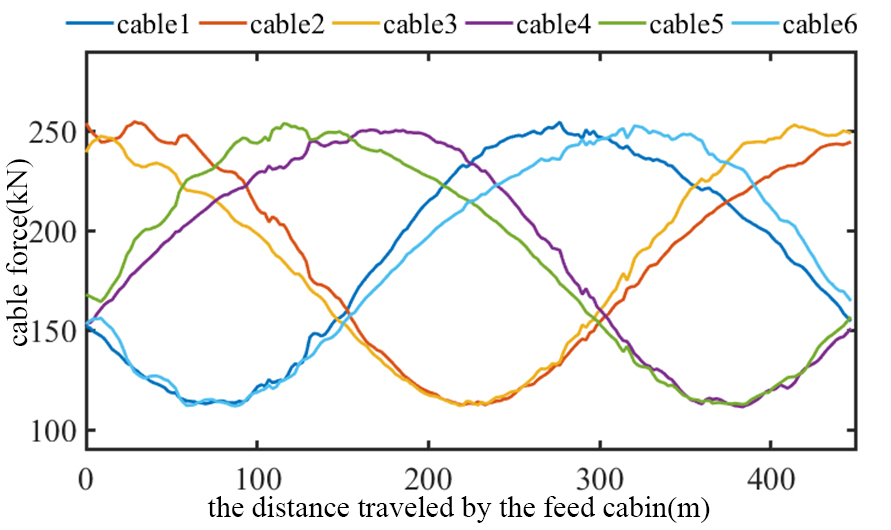}\\
		{\tiny (a) Cable forces solved by AAS with actual attitudes}
	\end{minipage}%
	\begin{minipage}[t]{0.495\textwidth}
		\centering
		\includegraphics[width=70mm]{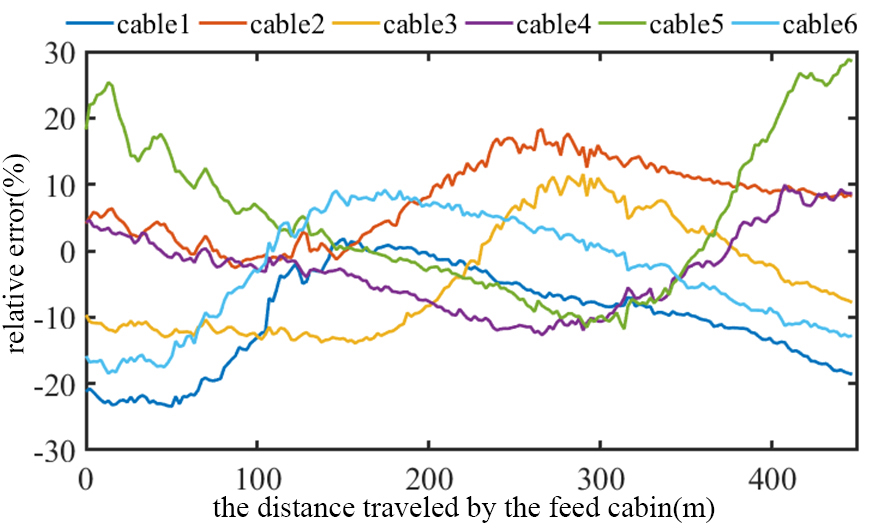}\\
		{\tiny (b) Relative errors between AAS and real-time control}
	\end{minipage}%
	\caption{\small Comparison between the AAS method with actual attitudes and the real-time control.}
	\label{Fig:13}
\end{figure}

Figur~14(a) shows the cable forces calculated by the AAS method 
after considering the measured attitude of the feed cabin and correcting the mass center coordinates, 
while Figure~14(b) shows the relative errors between these calculated cable forces 
and the actual measured cable forces in Figure~12(b). 
And the root mean square errors between them are shown in Table~6. 
It can be found that the theoretical and experimental numerical trends are very close, 
and the relative errors are already acceptable.
There are still many reasons for these errors.

\begin{figure}[h]
	\begin{minipage}[t]{0.495\linewidth}
		\centering
		\includegraphics[width=70mm]{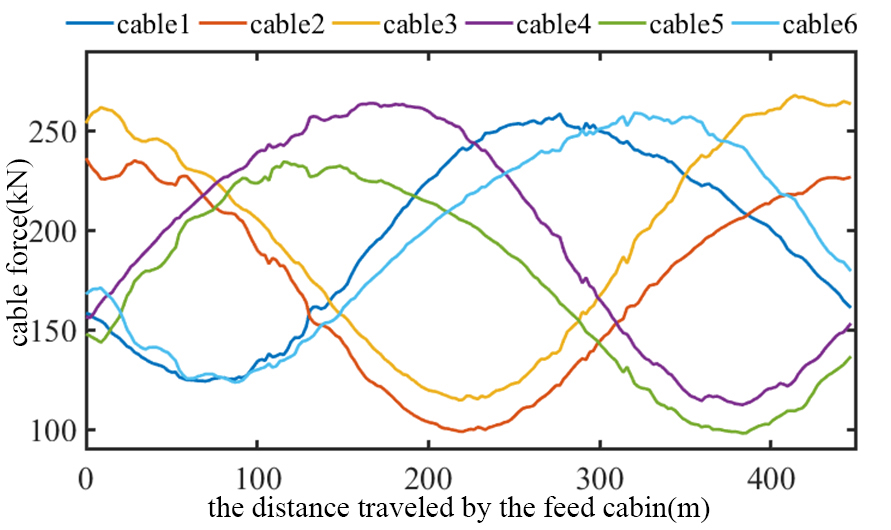}\\
		{\tiny (a) Cable forces solved by AAS with actual attitudes and corrected coordinates}
	\end{minipage}%
	\begin{minipage}[t]{0.495\textwidth}
		\centering
		\includegraphics[width=70mm]{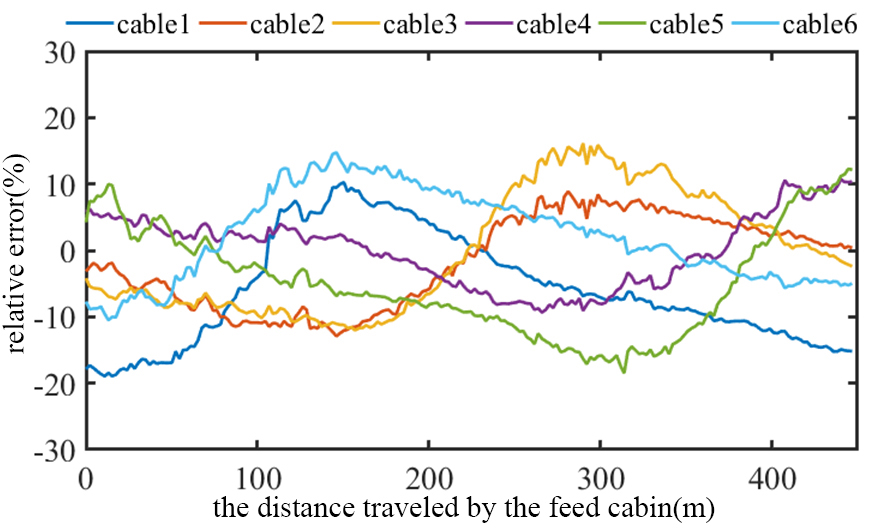}\\
		{\tiny (b) Relative errors between AAS and real-time control}
	\end{minipage}%
	\caption{\small Comparison between the AAS method with actual attitudes and corrected coordinates and the real-time control.}
	\label{Fig:14}
\end{figure}

The first is the coordinate deviation of the mass center. 
Even a slight error after correction can have a huge impact on the cable forces. 
Moreover, the structure of the feed cabin actually changes during operation, 
which also causes the change in the mass center coordinates more or less.

The second is the fact that many wires and sensors are added to the cables, 
which results that the cables are not of uniform quality assumed by theory. This causes the deviations in the cable forces.

Third, the object of comparison is the result of the existing model combined with PID control. 
Although it has been verified and can be used, it still has errors compared with true values.

In addition, the sensors on the cables also have a measurement error of about 3\%, 
which causes deviations in the cable forces as well.

In summary, the theoretical values calculated by AAS after considering the measured attitudes 
and the corrected mass center are consistent with the trend of the experimental data. 
It is expected to replace the existing model of FAST in the future 
and combine with PID control or even machine learning to achieve more efficient and accurate control.

\section{Conclusions}
\label{sect:Conclusions}

In this paper, a fast method, called AAS, 
for solving the static or quasi-static equations of the CDPR of FAST is proposed to achieve FAST’s real-time control. 
By extracting the necessary geometric and physical coefficients, the static or quasi-static equations can be solved analytically. 
In the comparison example with the traditional iterative method, 
AAS can save at least 75\% of the time in the calculation of single cables’ force at a certain moment 
and even can save 90\% of the time in the calculation of single cables’ force during the CDPR slowly moves. 
Also, it is verified through the experiment that the values calculated by AAS are consistent with the measured data.
Obviously, the difficulty of using the catenary model to control FAST in real-time is solved. 
Presumably in the future, 
FAST can be controlled with higher precision and more efficient to complete more and more difficult observation tasks, 
and this method may be extended to other CDPRs.

\begin{acknowledgements}
This work was financially supported by 
the National Natural Science Foundation of China (Nos. 11673039,11973062); 
the Youth Innova-tion Promotion Association CAS; 
the Open Project Program of the Key Laboratory of FAST, NAOC, Chinese Academy of Sciences.
\end{acknowledgements}

\label{lastpage}

\end{document}